\documentclass[showpacs,amssymb,floatfix]{revtex4}
\usepackage{graphicx}
\usepackage{dcolumn}
\usepackage{bm}
\begin{document}
\title{Symmetry breaking in binary chain with nonlinear sites}

\author{Dmitrii N. Maksimov and Almas F. Sadreev}
\affiliation{Institute of Physics, 660036, Krasnoyarsk, Russia}
\date{\today}
\begin{abstract}
We consider a system of two or four nonlinear
sites coupled with binary chain waveguides.
When a monochromatic wave is injected into the first
(symmetric) propagation channel the presence of cubic nonlinearity
can lead to symmetry breaking giving rise to emission of
antisymmetric wave into the second (antisymmetric) propagation
channel of the waveguides. We found that in the case of nonlinear
plaquette there is a domain in the parameter space where neither
symmetry preserving nor symmetry breaking stable stationary
solutions exit. As a result injection of a monochromatic symmetric
wave gives rise to emission of nonsymmetric satellite waves with
energies different from the energy of the incident wave. Thus, the response exhibits
nonmonochromatic behavior.
\end{abstract}
 \maketitle
\section{Introduction}

To the best our knowledge symmetry breaking (SB) in the nonlinear
 systems was first predicted by Akhmediev who considered a
composite structure of a single linear layer between two
symmetrically positioned nonlinear layers \cite{Akhmediev1}. One
could easily see that if the wave length is larger than the
thickness of the nonlinear layers then Akhmediev's model could be
reduced to a dimer governed by the nonlinear Shr\"odinger
equation. Independently the SB was discovered for the discrete
nonlinear Schr\"odinger equation with a finite number of coupled
cites (nonlinear dimer, trimer, {\it etc})
\cite{Eilbeck,Tsironis,Kenkre0,Kenkre,Bernstein,Tsironis1,Molina}.
For example, in the case of the nonlinear Schr\"odinger dimer
Eilbeck {\it et al} found two different families of stationary
solutions \cite{Eilbeck}. The first family is symmetric
(antisymmetric) ($ |\phi_1|= |\phi_2|$) while the second is
nonsymmetric ($|\phi_1|\neq |\phi_2|$). This consideration was
later extended to a nonlinear dimer embedded into an infinite
linear chain \cite{brazhnyi} with the same scenario for the SB.
Multiple bifurcations to the symmetry breaking solutions were
demonstrated by Wang {\it et al} \cite{Malomed4} for the nonlinear
Schr\"odinger equation with a square four-well potential.
Remarkably, the above system can also support a stable state with
a nodal point, i.e., quantum vortex \cite{Kevrekidis}. In the
framework of the nonlinear Schr\"odinger equation one can achieve
bifurcation to the states with broken symmetry varying the
chemical potential which is equivalent to the variation of the
population of the nonlinear sites or, analogously, of the
 constant in the nonlinear term of the Hamiltonian. In
practice, however, one would resort to the optical counterparts of
the quantum nonlinear systems where the variation of the amplitude
of the injected wave affects the strength of Kerr nonlinearity
(see Refs.
\cite{Yabuzaki,Otsuka,Haelterman,Babushkin,Kevrekidis1,Segev,Maes1,Audin,Li,BPS,T}
for optical examples of the SB).

In the present paper we consider a nonlinear dimer and a square
four-site nonlinear plaquette coupled with two channel waveguides
in the form of binary tight-binding chains. The latter system is
analogous to that recently considered in Ref. \cite{Lepri} where
the plaquette however was set up without mirror symmetry with
respect to the center-line of the waveguide. Due to nonlinearity
of the plaquette the system can spontaneously bifurcate between
diode-antidiode and bidirectional transmission regimes.
In our case, we will show that when a monochromatic wave is
injected into the first (symmetric) propagation channel the
presence of the cubic nonlinearity can lead to symmetry breaking
giving rise to emission of antisymmetric wave into the second
(antisymmetric) propagation channel of the waveguides.

This result raises an important question about the effect of a
probing wave on the SB. To allow for symmetry breaking the
architecture of the open system should support some symmetries of
its closed counterpart. When the growth of the injected power can
result in bifurcations into the states violating the symmetry of
the probing wave by the amplitude of the scattering function
\cite{Yabuzaki,Otsuka,Haelterman,Maes1,Maes2,Segev,Li,BPS} or by its phase \cite{JOSAB}.

The second important question is whether the solutions in the
linear waveguides could be stationary monochromatic plane waves,
reflected $\psi(n,t)=R\exp(-ikn-iE(k)t)$ and transmitted
$\psi(n,t)= T\exp(ikn-iE(k)t)$, where $R$ and $T$ are the
reflection and transmission amplitudes. If the answer is positive
then we can apply the Feshbach projection technique
\cite{Feshbach,Ingrid,Datta,Dittes,SR} and implement the formalism
of non-Hermitian effective Hamiltonian (now nonlinear) which acts
on the nonlinear sites only thus truncating the Hilbert space to
the scattering region \cite{SR,BPS}. When the radiation shifts of
the energy levels are neglected that formalism reduces to the well
known coupled mode theory (CMT) equations
\cite{haus,manol,fan-suh,suh}.  In
the present paper we will show that there are domains in the
parameter space where there are no stable stationary solutions. We
will demonstrate numerically that a plane wave incident onto a
nonlinear object gives rise to emission of multiple satellite
waves with energies (frequencies) different from the energy
(frequency) of the probing wave.


\section{Structure lay-out and basic equations}
\begin{figure}[ht]
\includegraphics[scale=0.6]{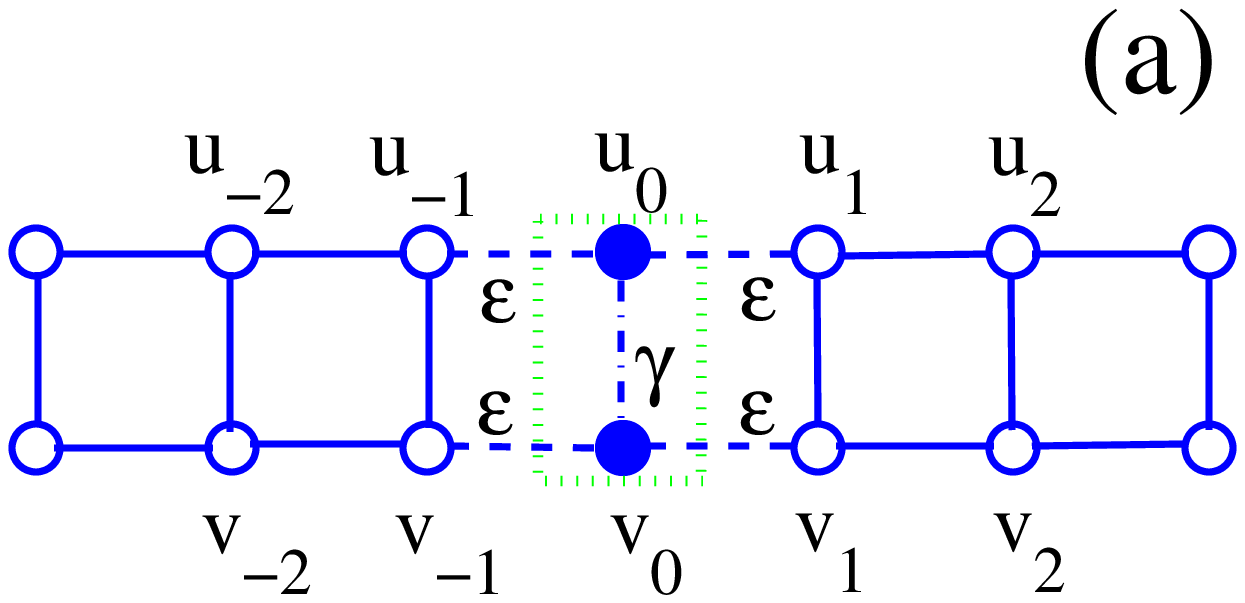}
\includegraphics[scale=0.6]{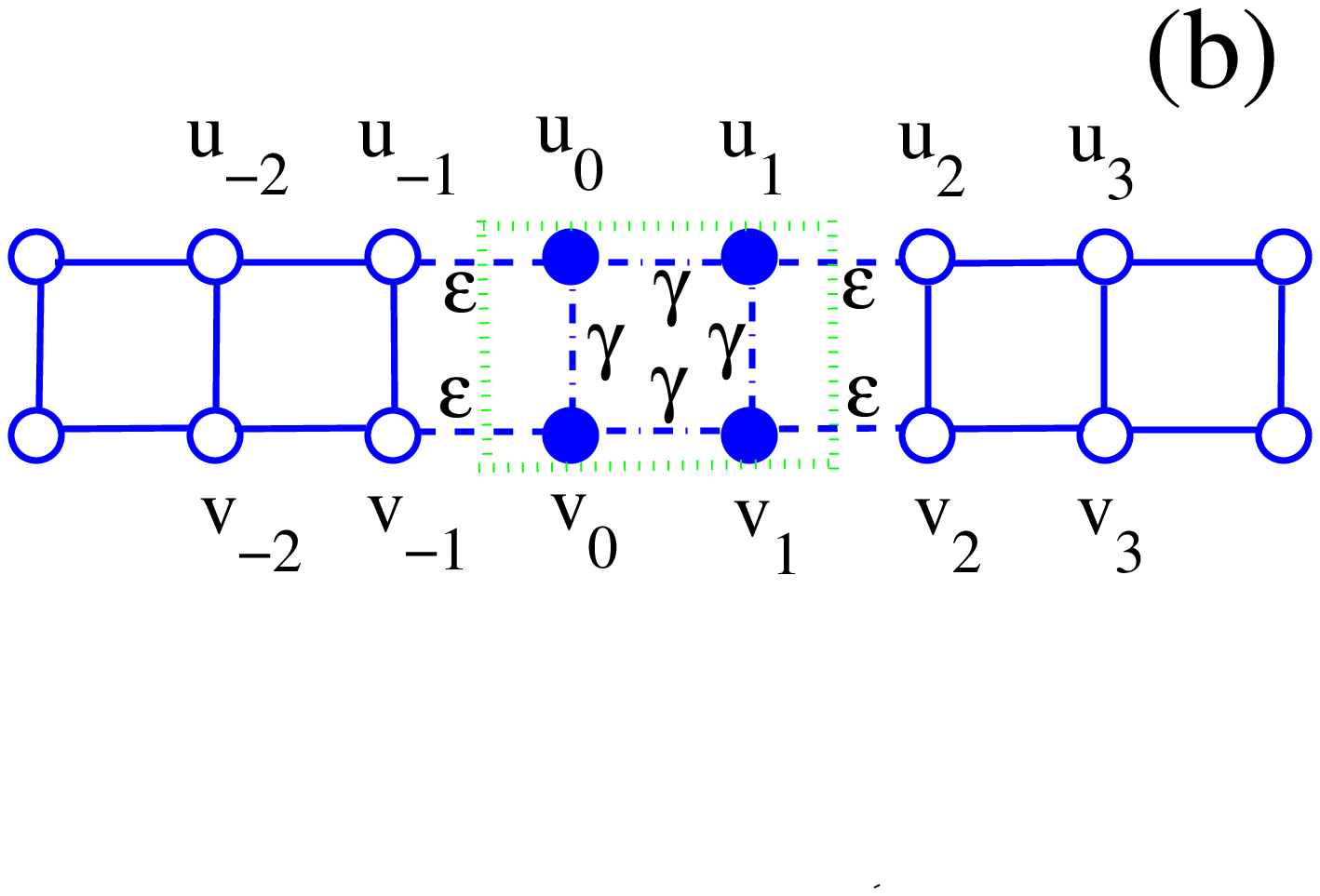}
\caption{(a) Two-channel waveguide in form of binary
tight-binding chain which holds a nonlinear dimer (shown by filled
circles). (b) The same as in subplot (a) but with four nonlinear sites (four-site plaquette). Green
dotted box is placed around the nonlinear scattering region.}
\label{fig1}
\end{figure}

We consider two tight-binding structures shown in Fig. \ref{fig1},
nonlinear dimer (a) and nonlinear four-site square plaquette (b)
coupled with linear binary chains. Each chain, left and right,
supports two continua of plane waves
\begin{equation}\label{cont}
    \psi_1^{(\pm)}(n,t)=\frac{1}{\sqrt{2|\sin k_1|}} \left(\begin{array}{c} 1\cr 1\end{array}\right)
    \exp(\pm ik_1n-iE(k_1)t), \ \
    \psi_2^{(\pm)}(n,t)=\frac{1}{\sqrt{2|\sin k_2|}} \left(\begin{array}{c} 1\cr - 1\end{array}\right)
    \exp(\pm ik_2n-iE(k_2)t)
\end{equation}
where indices $p=1, 2$ enumerate continua or channels with the
propagation bands given by
\begin{equation}\label{bands}
    E(k_p)=-2\cos k_p\mp 1, -\pi\leq k_p \leq \pi.
\end{equation}
In order to control the resonant transmission we adjust the
hopping matrix element between the nonlinear sites and waveguides
$\epsilon<1$ \cite{SR} shown in Fig. \ref{fig1} by dash lines. The
coupling between nonlinear sites (shown in Fig. \ref{fig1} by
dash-dot line) is controlled by constant $\gamma$.

In the case of the nonlinear dimer the Schr\"odinger equation
takes the following form

\begin{eqnarray}\label{SE}
i\dot{u}_n=-J_{n}u_{n+1}-J_{n-1}u_{n-1}-K_nv_{n}+
\lambda\delta_{n,0}|u_n|^2u_n,\nonumber\\
i\dot{v}_n=-J_{n}v_{n+1}-J_{n-1}v_{n-1}-K_nu_{n}+\lambda\delta_{n,0}|v_n|^2v_n,
\end{eqnarray}
where $J_{n}=1+(\epsilon-1)(\delta_{n,-1}+\delta_{n,0})$,
$K_{n}=1+(\gamma-1)\delta_{n,0}$. In the case of nonlinear square plaquette
the Shcr\"odinger equation could be written down essentially in
the same way.



\section{Nonlinear dimer}

First, let us follow Refs.\cite{mcgurn,flach,longhi,miros,ming2,miros1,miros2} and search for the solution of the Schr\"odinger equation
(\ref{SE}) in the form of a stationary
wave
\begin{equation}\label{stat}
    u_n(t)=u_ne^{-iEt}, v_n(t)=v_ne^{-iEt}
\end{equation}
where the discrete space variable $n$ and the time $t$ are separated. The absence of nonlinearity
in the waveguides drastically simplifies analysis of Eq.
(\ref{SE}). Assuming that a symmetric/antisymetric wave is incident from
the left we can write the solutions in the left $\psi_{L}(n,t)$ and right $\psi_{R}(n,t)$ waveguides as:
\begin{eqnarray}\label{plane}
\psi_{L}(n,t)=A_0\psi_p^{(+)}(n,t)+ R_{p,1}\psi_{1}^{(-)}(n,t)+R_{p,2}\psi_2^{(-)}(n,t), \nonumber \\
\psi_{R}(n,t)=T_{p,1}\psi_1^{(+)}(n,t)+T_{p,2}\psi_2^{(+)}(n,t),
\end{eqnarray}
where parameter $A_0$ is introduced to tune the intensity of the
probing wave. Notice that Eq. (\ref{plane}) implicitly defines
reflection and transmission amplitudes $R_{p,p'}$ and $T_{p,p'}$
with the first subscript $p$ indexing the channels and
respectively the symmetriy of the incident wave. One can now match
the solutions Eq. (\ref{plane}) to the equations for the nonlinear
sites to obtain a set of nonlinear equations for on-site and
reflection/transmission amplitudes. Computationally, however, it
is more convenient to use the approach of the non-hermitian
Hamiltonian \cite{Datta,SR} in which the number of unknown
variables equals to the number of nonlinear sites. Following Ref.
\cite{BPS} we write the equation for the amplitudes on the
nonlinear sites
\begin{equation}\label{LS}
    (E-H_{eff})|\psi\rangle=A_0\epsilon|in\rangle
\end{equation}
where
\begin{equation} \label{Datta}
H_{eff}=H_0-\sum_CV_C^{\dagger}\frac{1}{E+i0-H_C}V_C=
\left(\begin{array}{cc}-\epsilon^2 (e^{ik_1}+e^{ik_2})+\lambda|u_0|^2 & -\gamma - \epsilon^2 (e^{ik_1}-e^{ik_2})\cr
 -\gamma - \epsilon^2 (e^{ik_1}-e^{ik_2}) & \epsilon^2 (e^{ik_1}+e^{ik_2})+\lambda|v_0|^2
\end{array}\right).
\end{equation}
Here vector $\langle \psi| = (u_0, v_0)$ is the state vector of
the dimer, $H_0$ is the nonlinear Hamiltonian of the dimer
decoupled from the waveguides, $V_C$ is the coupling operator
\cite{SR} between the nonlinear sites and the left and right
waveguides $C=L,R$ with the Hamiltonian $H_C$, and $\langle
in|=(1, \pm 1)$ is the source term for symmetric/antisymmetric
incident waves correspondingly. One can easily see that in the
limit $\epsilon \rightarrow 0$ the dimer is decoupled from the
waveguides and Eq. (\ref{LS}) limits to the standard nonlinear
Schr\"odiner equation of the closed nonlinear dimer. After Eq.
(\ref{LS}) is solved one easily obtains transmission/reflection
amplitudes from Eq. (\ref{SE}).

\begin{figure}
\includegraphics[height=6cm,width=7cm,clip=]{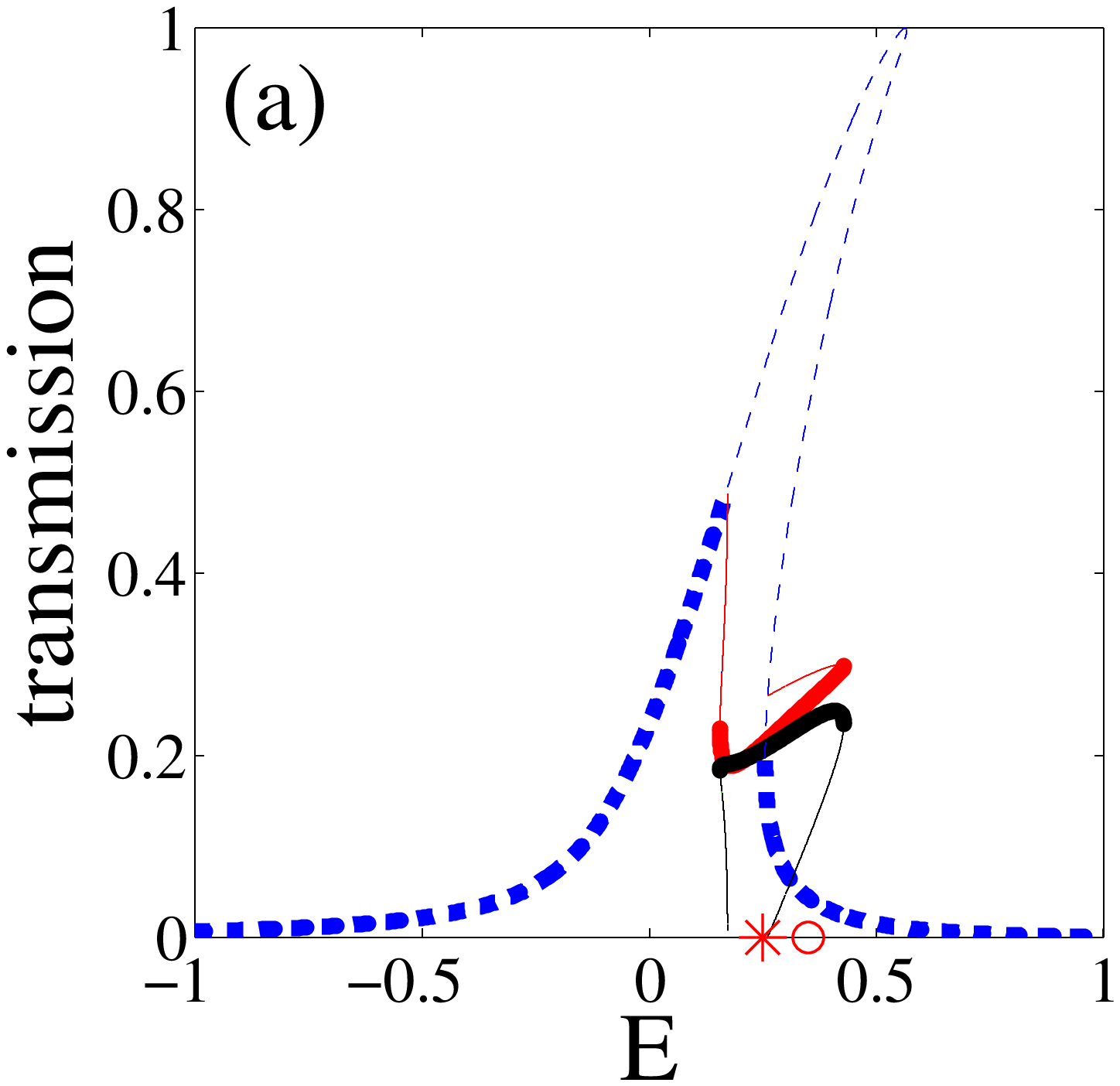}
\includegraphics[height=6cm,width=7cm,clip=]{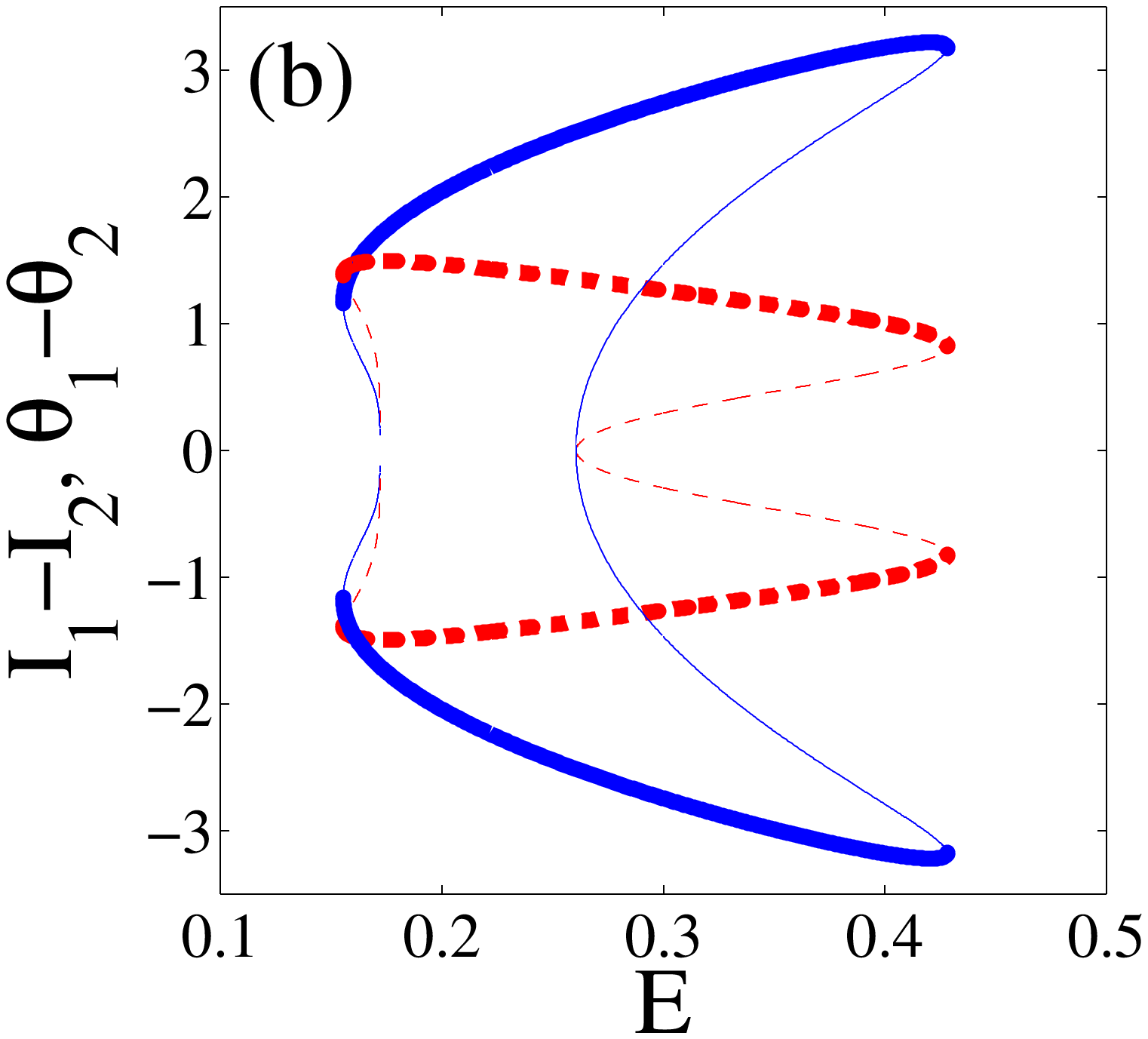}
\caption{(Color online) Response of nonlinear dimer to symmetric probing wave.
(a) Transmission probability $|T_{11}|^2$ for symmetry preserving solution -
blue dash line, $|T_{11}|^2$ for SB solutions - solid red line,
$|T_{12}|^2$ SB solutions - solid black. (b) Phase difference
$\theta_2-\theta_1$ for the pairs of SB solutions - dashed red line; intensity difference $I_1-I_2$ - solid blue line.
Thicker lines mark stable solutions. The parameters: $\epsilon=0.2, A_0=1, \lambda=0.025, \gamma=0$. }
\label{fig2}
\end{figure}

In Fig. \ref{fig2} we present the response of the nonlinear dimer to a symmetric probing wave both in form of
transmission probabilities $|T_{11}(E)|^2$, $|T_{12}(E)|^2$ Fig. \ref{fig2}(a), and
phase $\theta_2-\theta_1$ and intensity $I_1-I_2$ differences Fig. \ref{fig2}(b). On-site phases and intensities
are defined through $u_0=\sqrt{I_1}e^{i\theta_1}, v_0=\sqrt{I_2}e^{i\theta_2}$. There are two principally different
families of solutions similar to those obtained in Ref. \cite{BPS}. The
symmetry preserving family inherits the features of the linear case in which
a symmetric probing wave excites only symmetric mode $\phi_s=\frac{1}{\sqrt{2}}(1, 1)$
with eigenvalue $E_s=-\gamma$. In fact, we see a typical example of
nonlinear transmission through a resonant state. The transmission probability $|T_{11}(E)|^2$ for
the symmetry preserving family of solutions with $I_1=I_2, \theta_1=\theta_2$ are shown
in Fig. \ref{fig2} (a) by blue dashed lines. The energy behavior of the
transmission is very similar to the case of a linear waveguide
coupled with one nonlinear in-channel site
\cite{mcgurn,flach,yanik}. However, one can see that the peak of the resonant transmission is now
unstable which is a key to mode conversion as it will be shown below. The stability of the
solutions was examined by standard methods based on small
perturbations technique (see, for example,
Refs. \cite{Litchinitser,Cowan}).

\begin{figure}
\includegraphics[height=6cm,width=7cm,clip=]{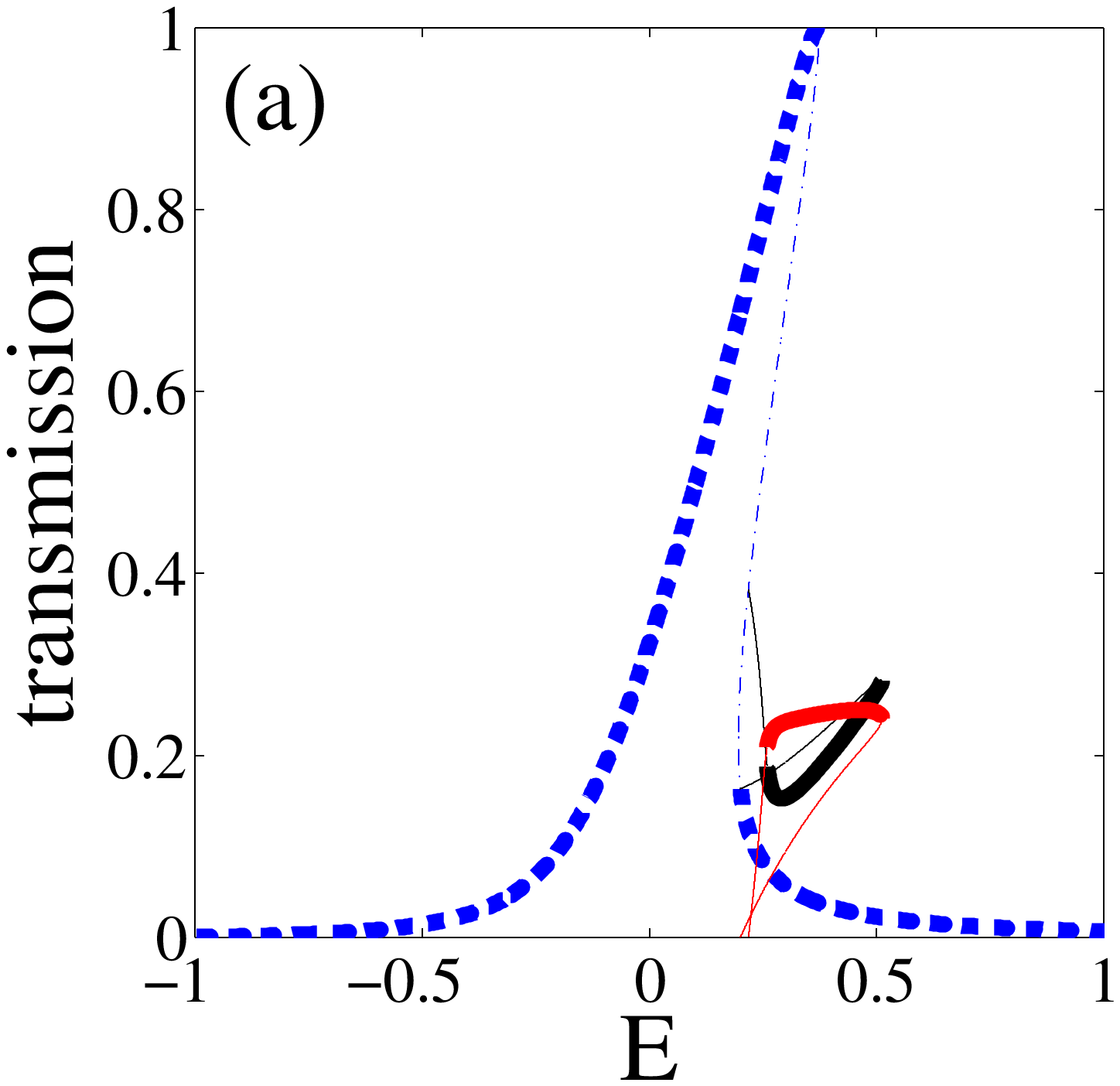}
\includegraphics[height=6cm,width=7cm,clip=]{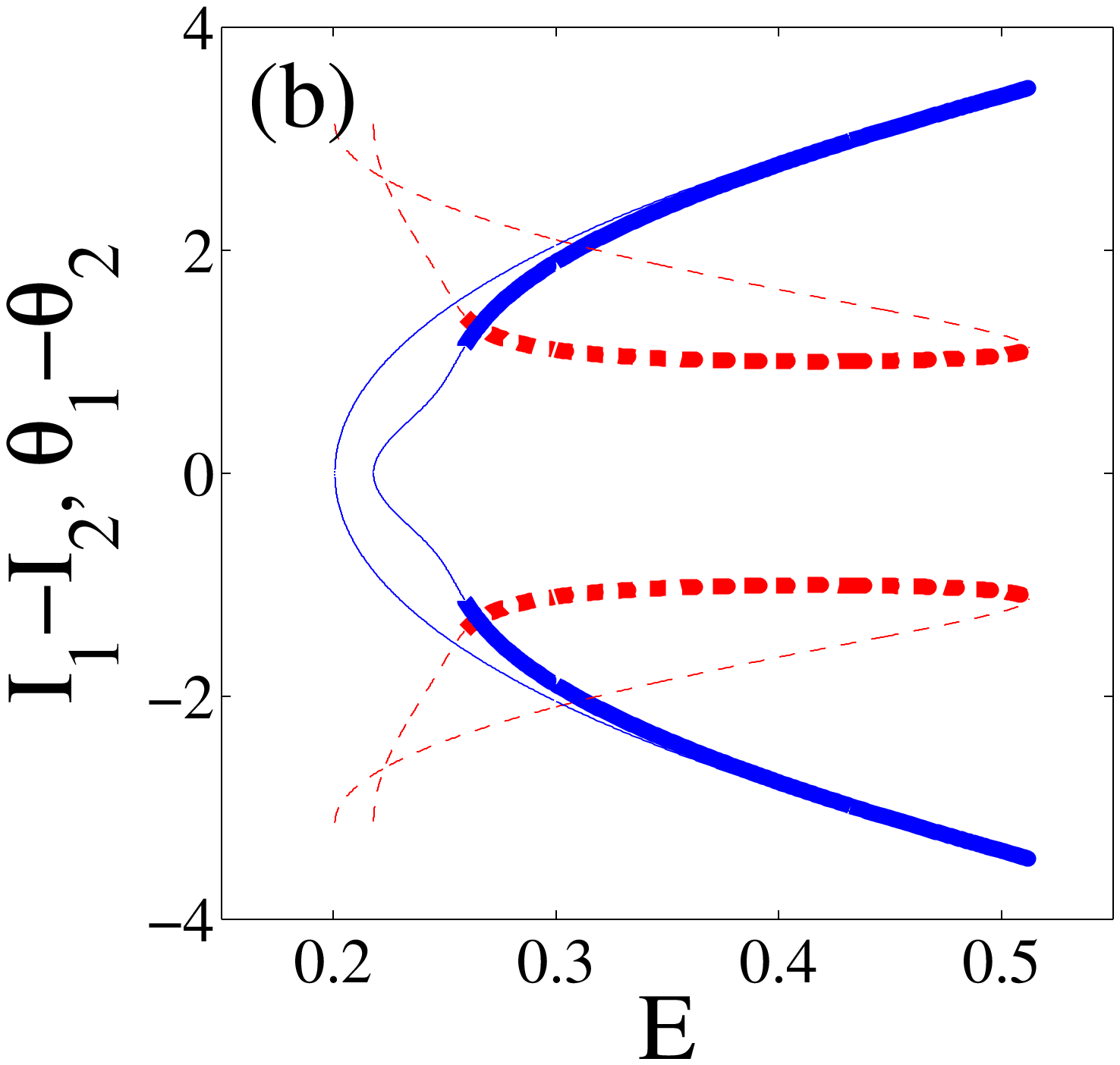}
\caption{(Color online) Response of nonlinear dimer to antisymmetric probing wave. (a) Transmission
probability $|T_{22}|^2$ for the symmetry preserving solution - blue dash line; $|T_{22}|^2$ for SB solutions -
solid red line,
$|T_{21}|^2$ SB solutions - solid black line. (b) Phase difference
$\theta_2-\theta_1$ for the pairs of SB solutions - dashed red lines, intensity difference $I_1-I_2$ - solid blue lines.
Thicker lines mark stable solutions. The parameters: $\epsilon=0.2, A_0=1, \lambda=0.03, \gamma=0$. } \label{fig3}
\end{figure}

The SB occurs due to excitation of the second (antisymmetric) mode of the dimer $\phi_a=\frac{1}{\sqrt{2}}(1,-1)$
with eigenvalue $E_a=\gamma$. The phase and intensity differences for the SB solutions are shown in Fig. \ref{fig2}(b).
One can see that the symmetry is broken simultaneously by both
intensity $I_1\neq I_2$ and phase $\theta_1 \neq
\theta_2$ of the wave function. This
scenario of SB differs from the case of the closed
dimer \cite{Eilbeck,Tsironis,Kenkre} or the dimer opened to a single chain (off-channel architecture)
\cite{BPS} where the symmetry was broken either by the
intensities or by the phases of the on-site amplitudes.

\begin{figure}
\includegraphics[height=6cm,width=7cm,clip=]{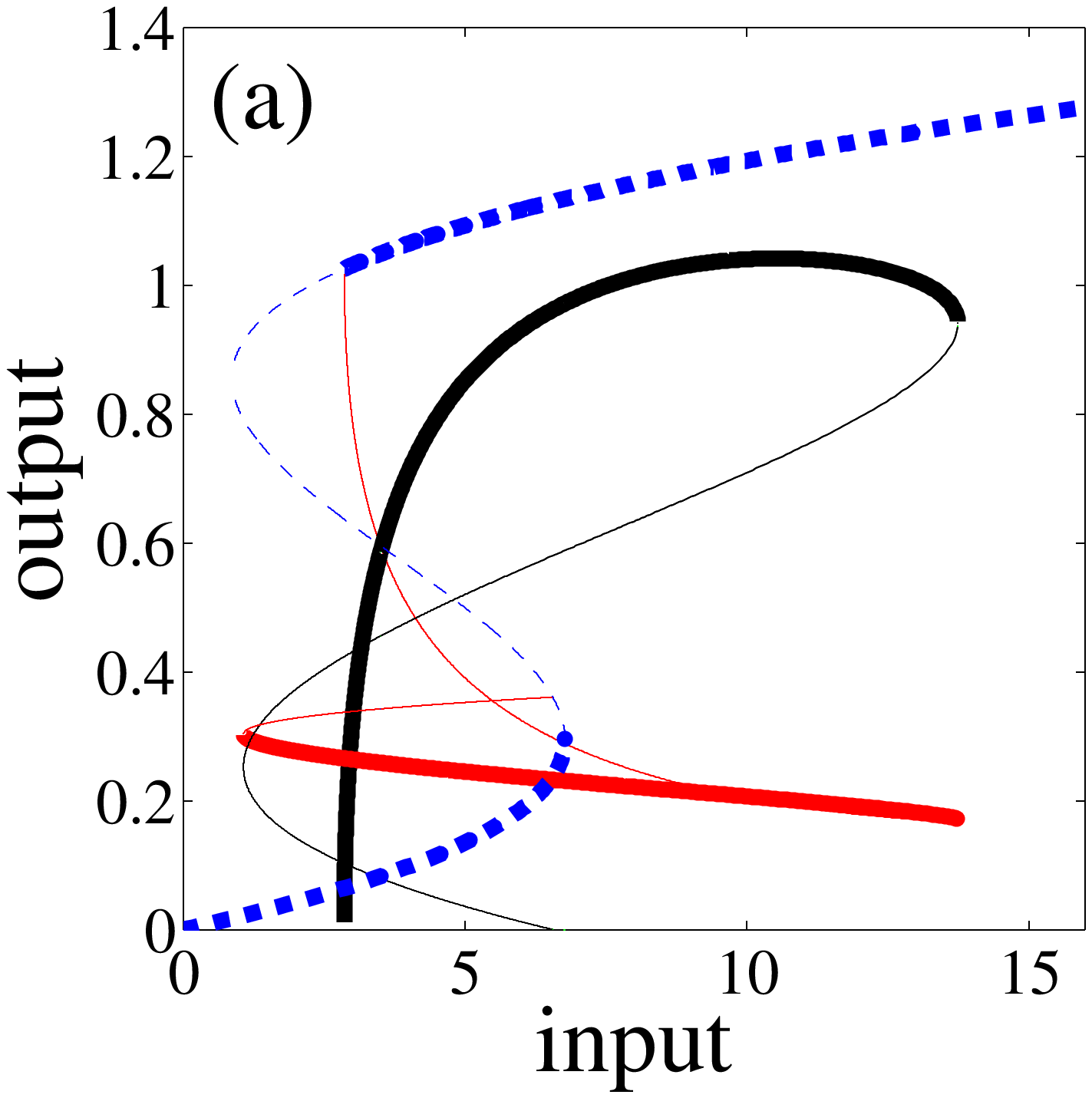}
\includegraphics[height=6cm,width=7cm,clip=]{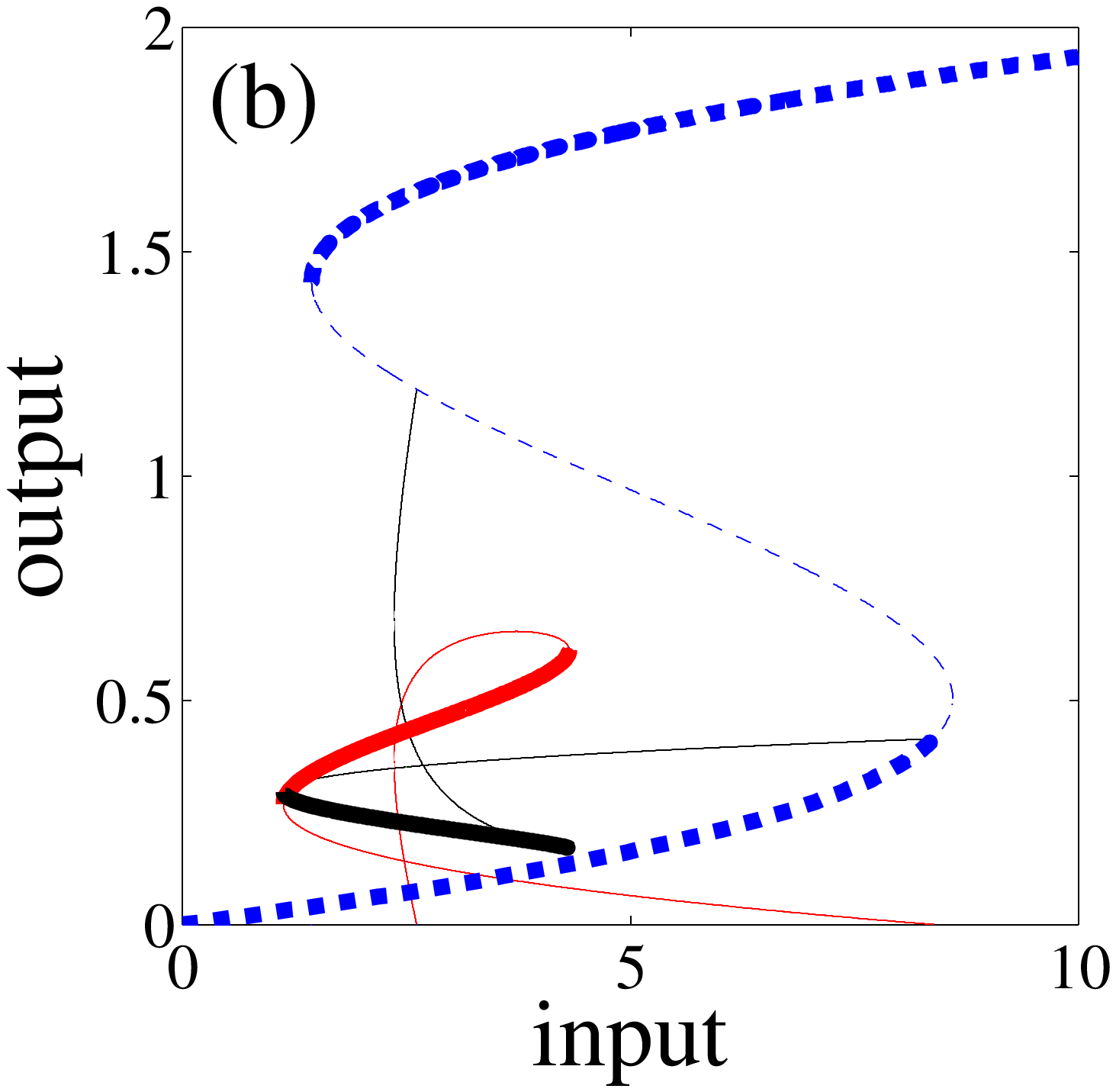}
\caption{(Color online) Input $A_0^2$ vs. outputs $|T_{p,p'}|^2$ for nonlinear dimer. (a) Symmetric probing
wave $\langle in|=(1, 1)$, $|T_{11}|^2$ - blue dash line
symmetry preserving solutions,  $|T_{11}|^2$ SB solutions - solid red line,
$|T_{12}|^2$ SB solutions - solid black line. (b) Antisymmetric probing
wave $\langle in|=(1, -1)$, $|T_{22}|^2$ - blue dash line
symmetry preserving solutions, $|T_{22}|^2$ SB solutions - solid red line,
$|T_{21}|^2$ SB solutions - solid black. The parameters: $\epsilon=0.2, E=0.45, \lambda=0.025, \gamma=0$. }
\label{fig4}
\end{figure}

In Fig. \ref{fig3}  we show the case when the antisymmetric wave
is injected to the system. Although the energy behavior of the
on-site intensities, on-site phases, and the transmission
probabilities to the first and second
channels $|T_{21}|^2, |T_{22}|^2$ is similar to the case of
symmetric probing wave, the stability pattern of the symmetry
preserving solution is now typical for the nonlinear transmission
through a single nonlinear site Ref. \cite{yanik}. Finally, following
Ref. \cite{joanbook} we present the plots output vs.
input, i.e., transmission probabilities $|T_{p,p'}|^2$ vs. $A_0^2$
for both symmetric and antisymmetric probing waves. In Fig. \ref{fig4} one can see
that the input-output curves for the symmetry preserving family of solutions
have typical bifurcation behavior while the symmetry breaking
family exists only within a bounded domain of the input.

As was mentioned above the SB phenomenon results in the mode
conversion, i.e., emission of scattered waves into a different channel. Let us consider time evolution of the
wave front given by
\begin{equation}\label{front}
    |\psi_{in}(n)\rangle=f(n)\psi^{(+)}_1(n,0),
\end{equation}
where
\begin{equation}
f(n)= \left\{
\begin{array}{cc} 1 & \mbox{if $n \leq -100$ }; \\ e^{-(n+100)^2/250}  & \mbox{if $n > -100$},
\end{array}
\right.
\end{equation}
is an auxiliary function that provides a smooth increase of
incident plane wave amplitude. As the wave front propagates to the
right the input signal converges to $\psi^{(+)}_1(n,t)$ so one
could expect that in course of time the system would stabilize in
one of the solutions shown in Fig. \ref{fig2}. In order to perform
numerical computations we directly applied both standard forth
order Runge-Kutta method and  the Besse-Crank-Nicolson relaxation
scheme \cite{Besse} to Eq. (\ref{SE}) with both techniques
yielding the same result (Crank-Nicolson approach being slightly
advantageous in terms of computational time). The absorbing
boundary conditions were enforced at far ends of the waveguides
according to Ref. \cite{Oskooi} to truncate the problem to a
finite domain.

\begin{figure}
\includegraphics[height=6cm,width=7cm,clip=]{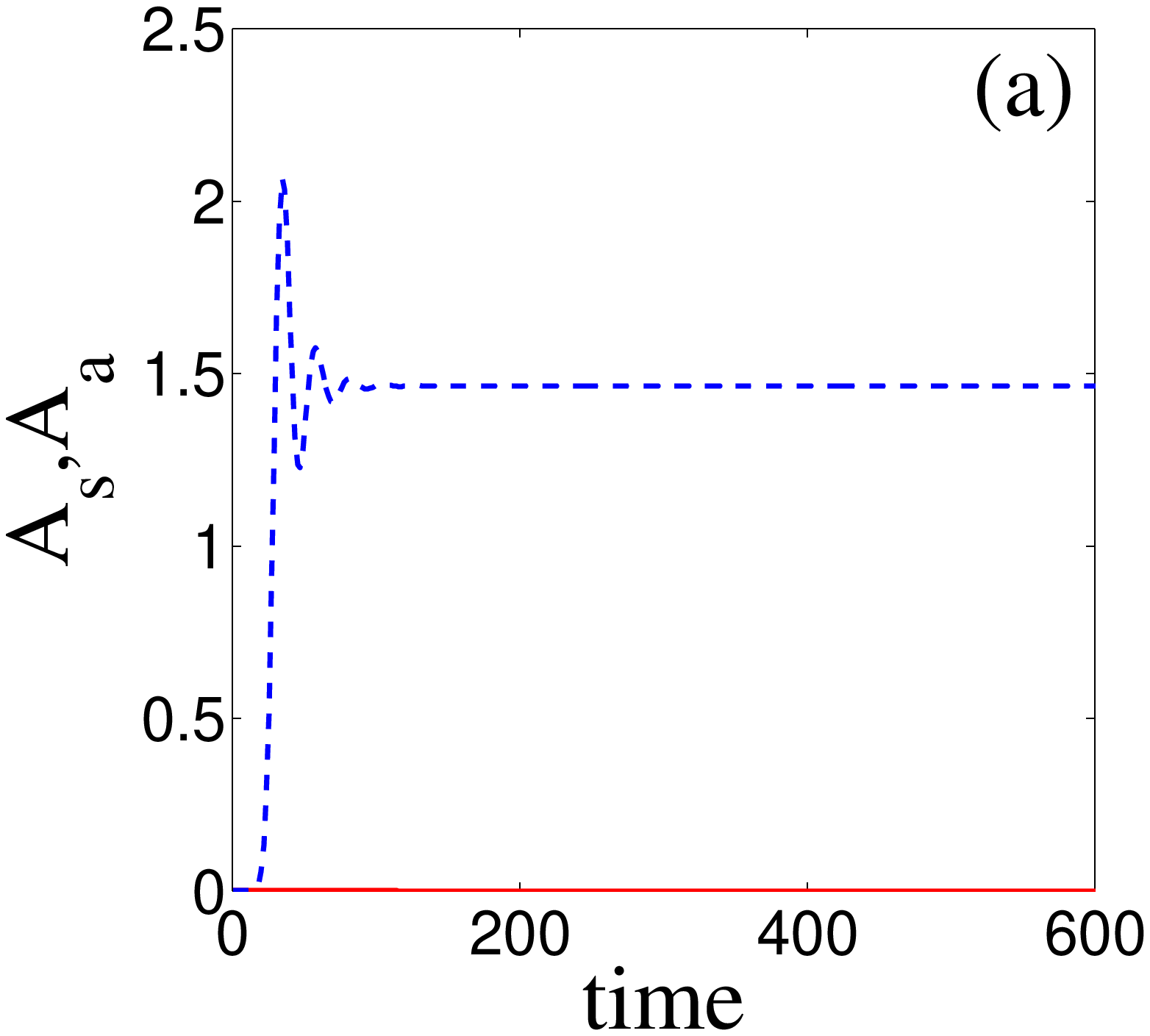}
\includegraphics[height=6cm,width=7cm,clip=]{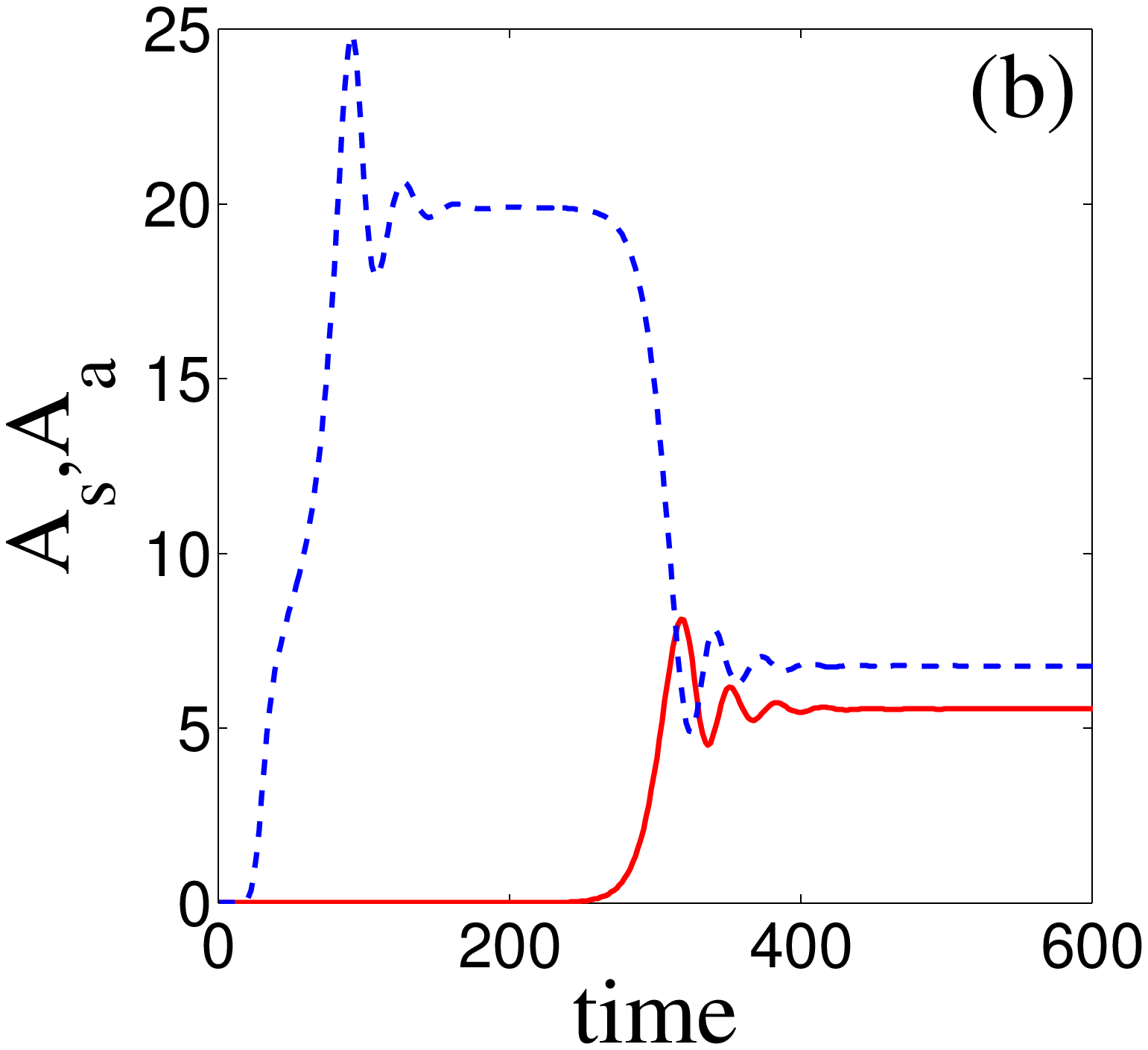}
\caption{(Color online) Time evolution of populations $A_s$ - dashed blue line, $A_a$ - solid red line. $\epsilon=0.2, A=1, \lambda=0.025, \gamma=0$. (a) Stable symmetry preserving solution
$E=0.35$ (red circle in Fig. \ref{fig2}(a)).
(b) Unstable symmetry preserving solution $E=0.25$ (red star in Fig. \ref{fig1}(a)). The parameters: $\epsilon=0.2, A_0=1, \lambda=0.025, \gamma=0$} \label{fig5}
\end{figure}

Let us first choose the energy in the domain where the symmetry
preserving solution is stable, for example $E=0.35$ (shown in Fig.
\ref{fig2}(a) by red circle). To detect the SB we use the populations $A_s$ and
$A_a$ of the symmetric $\phi_s$ and antisymmetric $\phi_a$ modes
correspondingly. Fig. \ref{fig5}(a) shows that although stable SB
solutions exit at $E=0.35$ the response to the probing signal Eq.
(\ref{front}) contains only symmetric contribution and after
several oscillations converges to the stationary transmission. On
the other hand, if one chooses the energy of the incident wave
from the domain  where the symmetry preserving solutions are
unstable $E=0.25$ (shown in Fig. \ref{fig2}(a) by red star), the system first stabilizes in an unstable
symmetry preserving solution but then, due to accumulation of
numerical round-off error, it is forced to leave the unstable
equilibrium and occupy one of the stable states with broken
symmetry Fig. \ref{fig3}(a). Thus, the round-off error induces the
SB playing the role of noise in a real experiment.
It is important to notice that the results in Fig. \ref{fig5}
agree with the date obtained with the use of the effective
non-Hermitian Hamiltonian Eq. (\ref{LS}).


\section{Four-site nonlinear plaquette}

In the previous section we demonstrated how the phenomenon of SB
occurs in the nonlinear dimer due to resonant excitation of both
symmetric and antisymmetric modes $\phi_s, \phi_a$. Notice that a
necessary condition for the SB is that both modes be near
degenerate to be excited at the same energy. This forced us to
chose the inner coupling constant $\gamma=0$. In fact, our
numerical tests show that the SB will quickly vanish as $\gamma$
increased. In that sense the set-up of four-site plaquette seems
more feasible because now the closed system supports two
degenerate modes of different symmetries at any value of $\gamma$,
namely $\langle \chi_s|=\frac{1}{2} (1,1,-1,-1)$ and $\langle
\chi_a|=\frac{1}{2}(1,-1,1,-1)$. The corresponding degenerate
eigenvalue is $E_{s,a}=0$.

\begin{figure}
\includegraphics[height=6cm,width=7cm,clip=]{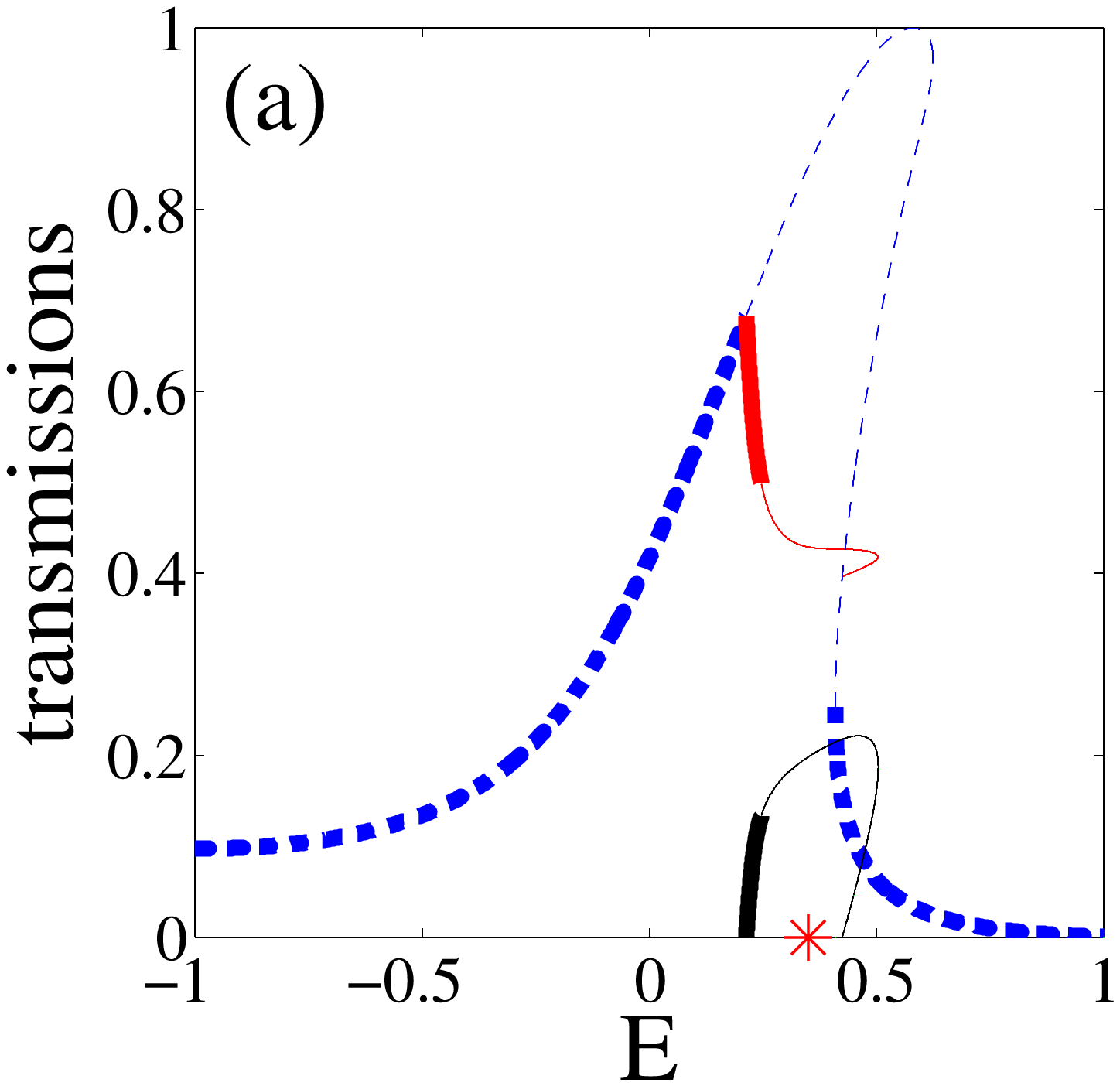}
\includegraphics[height=6cm,width=7cm,clip=]{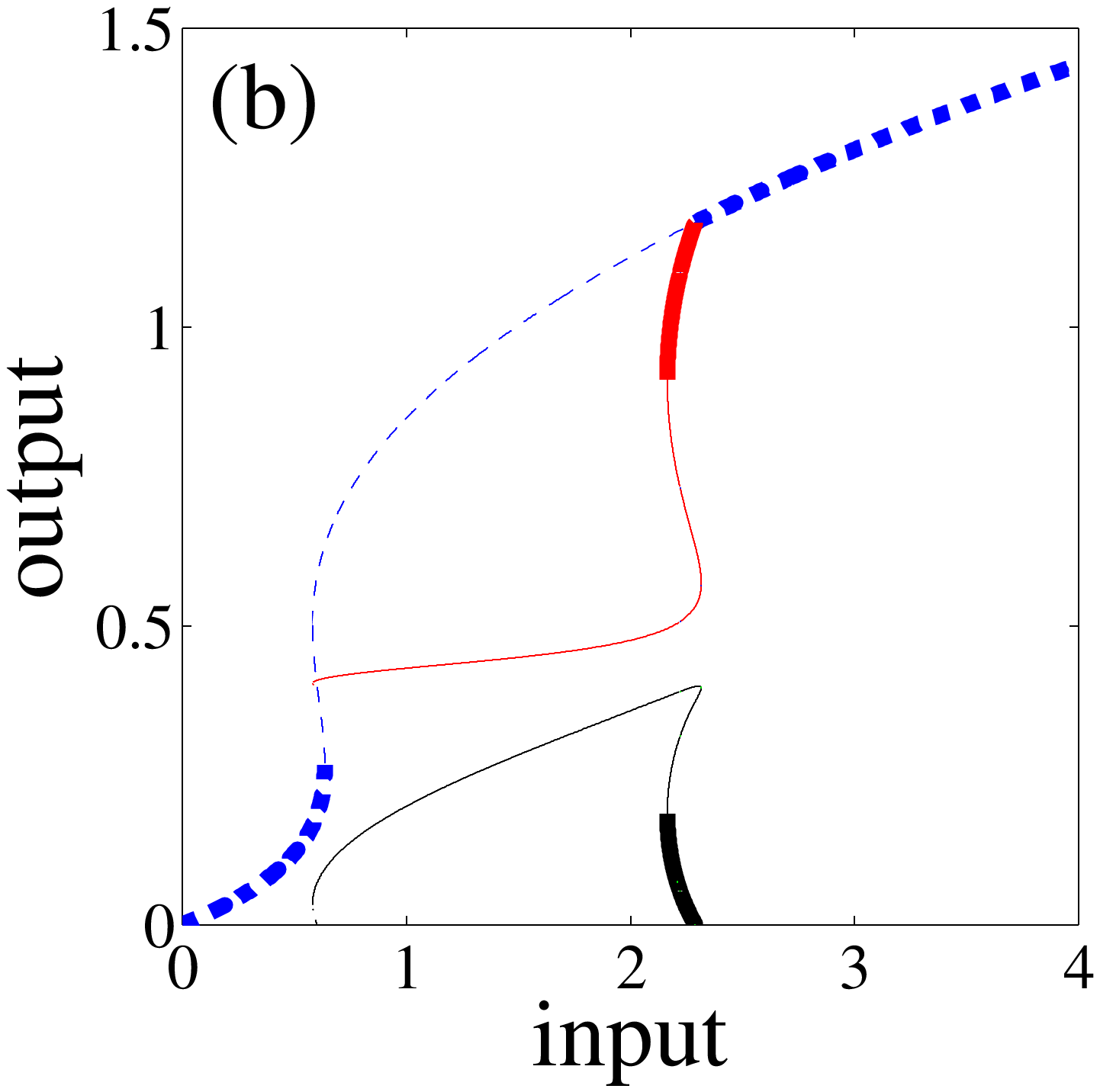}
\caption{(Color online) Response of nonlinear plaquette to symmetric probing wave $\epsilon=0.4, \gamma=1,
\lambda=0.09$, transmission probability $|T_{11}|^2$ for
symmetry preserving solutions - dashed blue line, $|T_{11}|^2$ SB solutions - solid red line,
$|T_{12}|^2$ SB solutions - solid black. (a) Transmissions vs. energy, $A=1$. (b) Transmissions vs. $A^2$, $E=0.35$.
Thicker lines mark stable solutions. In both cases one can see windows were neither symmetry preserving nor
symmetry breaking stable solutions exist.
}
\label{fig6}
\end{figure}

In the case of the four-site plaquette the effective Hamiltonian
takes the following form \cite{Datta,SR}
\begin{equation}\label{Datta4}
H_{eff}=\left(\begin{array}{cccc} -\epsilon^2 (e^{ik_1}+e^{ik_2})/2 +\lambda|u_0|^2
& -\gamma - \epsilon^2 (e^{ik_1}-e^{ik_2})/2 & -\gamma & 0\cr -\gamma - \epsilon^2 (e^{ik_1}-e^{ik_2})/2
& -\epsilon^2 (e^{ik_1}+e^{ik_2})/2 +\lambda|v_0|^2& 0 & -\gamma\cr
-\gamma& 0 & -\epsilon^2 (e^{ik_1}+e^{ik_2})/2 +\lambda|u_1|^2   &  -\gamma - \epsilon^2 (e^{ik_1}-e^{ik_2})/2 \cr
0& -\gamma &  -\gamma - \epsilon^2 (e^{ik_1}-e^{ik_2})/2  & -\epsilon^2 (e^{ik_1}+e^{ik_2})/2 +\lambda|v_1|^2
\end{array}\right).
\end{equation}
Eq. (\ref{LS}) should now be solved for the state vector of the
plaquette $\langle \psi| = (u_0, v_0, u_1, v_1)$ (see Fig.
\ref{fig1}(b)) with the source term $\langle in| = (1, \pm1, 0, 0)$.
The results of numerical solution are presented in Fig.
\ref{fig6}.

Although the result is similar to the case of nonlinear dimer (Figs.
\ref{fig2} and \ref{fig3}) there is one important difference.
Namely, there is now a domain the parameter space where {\it all}
stationary solutions are unstable as shown in Figs. \ref{fig6}.
Respectively, the solution of the transmission problem in such
system can be described neither by transmission and reflection
amplitudes Eqs. (\ref{plane}) nor by the Feshbach projection
method, i.e. by the the effective non-Hermitian Hamiltonian Eq.
(\ref{LS}). The problem of plane wave scattering from the
nonlinear plaquette can only be solved through numerical
simulation of the time-dependent equation.

\begin{figure}
\includegraphics[height=.45\textwidth,width=0.9\textwidth,clip=]{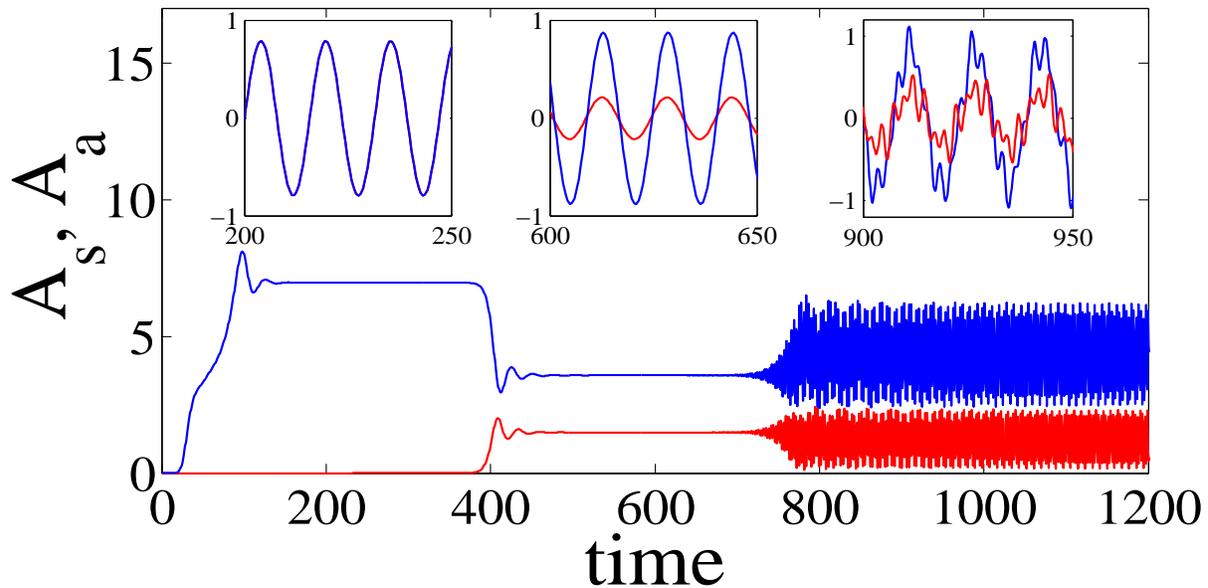}
\caption{(Color online) Time evolution  of populations of symmetric $B_s$ (blue line), and antisymmetric
$B_a$ (red line)  resonant modes,
 $\epsilon=0.4, \gamma=1, \lambda=0.09 E=0.4$.
The insets show the real parts of amplitudes $u_3$ (blue line), and $v_3$ (red line) vs. time $t$ in the corresponding regimes.
One can clearly see that in
course of time the system evolves to a non-stationary solution.} \label{fig7}
\end{figure}

\begin{figure}
\includegraphics[height=5cm,width=5cm,clip=]{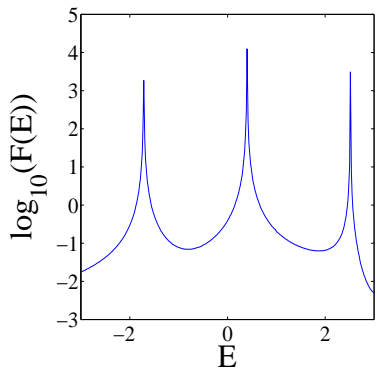}
\caption{(Color online) Logarithmic plot of Fourier power spectrum of output amplitude $u_3$ in the non-stationary regime Fig. \ref{fig7}.} \label{fig8}
\end{figure}

To perform numerical tests we repeated the wave-front simulations explained in the previous section.
Fig. \ref{fig7} shows the time evolution
 of populations $A_s, A_a$ of symmetric and antysimmetric resonant modes $| \chi_s \rangle, | \chi_a \rangle$.
At $t=0$ a symmetric wave front
(\ref{front}) is sent from the left waveguide towards the plaquette with its energy and amplitude
within the domain of unstable stationary
solutions $E=0.4$ (shown by red star in Fig \ref{fig6}(a)). First when $t\in [150,350]$ only symmetric state
of the plaquette is excited; $A_a=0$.
However, the symmetry preserving solution is unstable.
That causes transition to the symmetry breaking solution $A_a>0$ at
$t\approx 400$. As a result
the plaquette emits stationary plane waves of both symmetries with
the same energy as the probing wave when $t\in [450,650]$. Since this
solution is also unstable the system transits to another regime at $t\approx 700$. It is clearly seen in Fig. \ref{fig7}
that in this regime the solution is also symmetry breaking, however, what is more
interesting, it is {\it non}-stationary. The Fourier power spectrum $F(E)$ Fig. \ref{fig8} of the
amplitude $u_3$ (see Fig. \ref{fig1}) for $t\geq 800$ clearly shows
the presence of three peaks, the central peak with energy $E=0.4$ and two
satellites with energies $E_1=-1.71$, $E_2=2.51$.

With the growth of the nonlinearity constant $\lambda$ which is equivalent
to growth of the amplitude of the injected wave the dynamical properties of the nonlinear plaquette
change drastically.
\begin{figure}
\includegraphics[height=6cm,width=7cm,clip=]{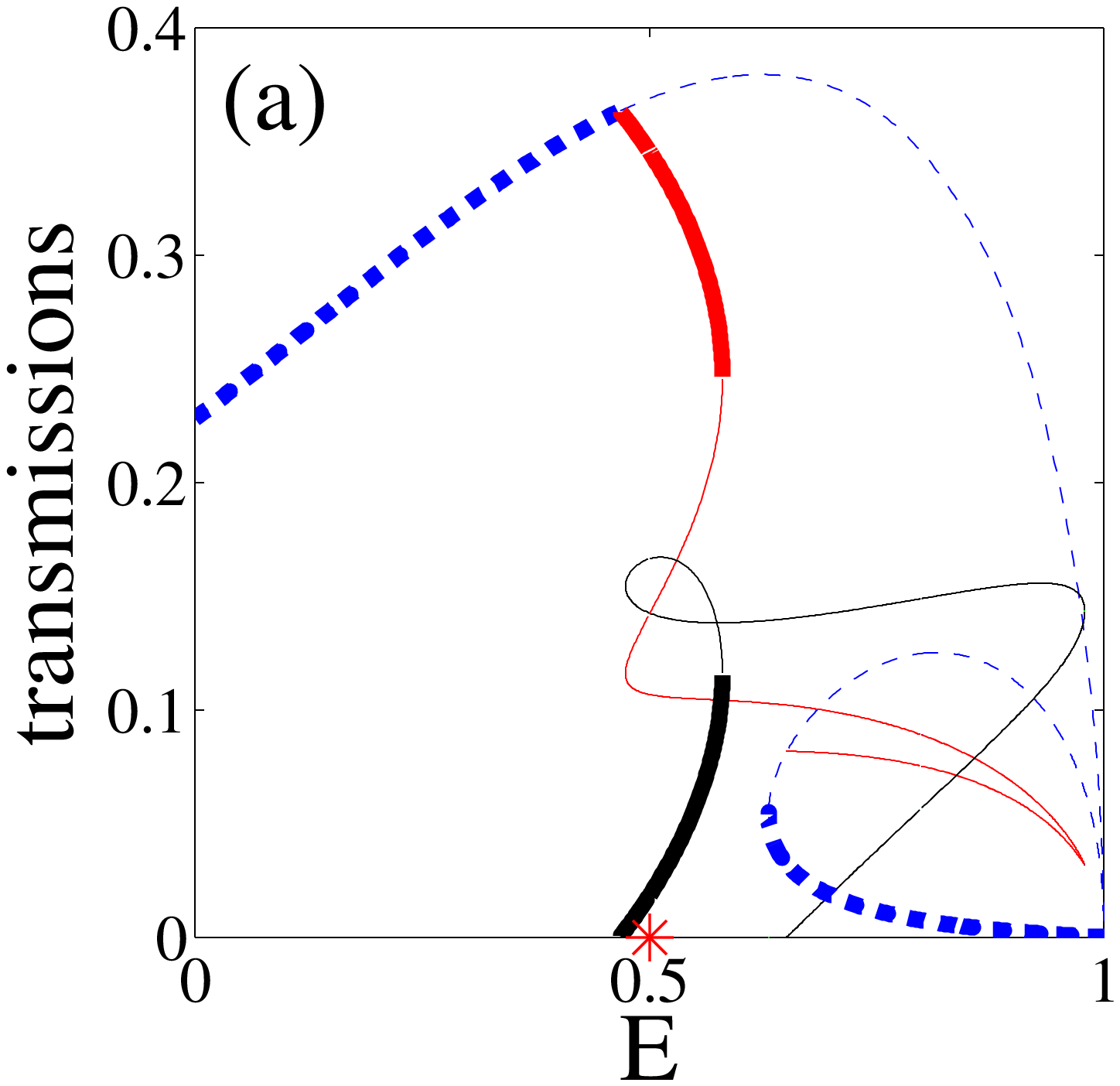}
\includegraphics[height=6cm,width=7cm,clip=]{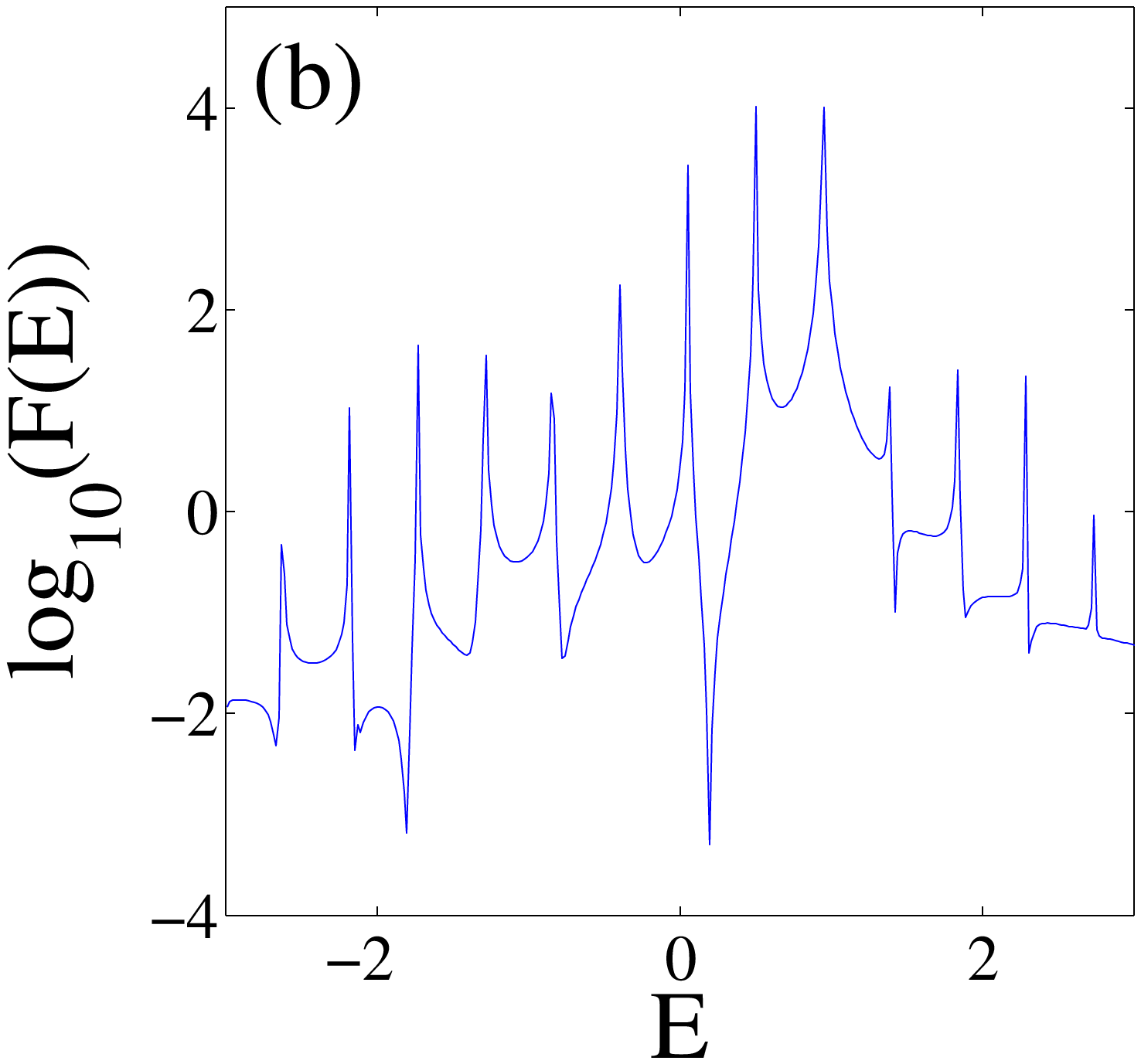}
\caption{(Color online) (a) Response of nonlinear plaquette to symmetric probing wave $\epsilon=0.4, \gamma=1,
\lambda=0.4, A_0=1$. Transmission probability $|T_{11}|^2$ for
symmetry preserving solutions - dashed blue line, $|T_{11}|^2$ SB solutions - solid red line,
$|T_{12}|^2$ SB solutions - solid black. (b) Corresponding Fourier power spectrum of amplitude $u_3$ at  $E=0.5$
(red star in Fig. \ref{fig9}(a)).} \label{fig9}
\end{figure}
Fig. \ref{fig9}(a) shows the
transmission probabilities vs. the incident energy. One can see that compared against Fig. \ref{fig6} the transmission peak
is now shifted towards the edge of the propagation band. When the energy of the incident wave  belongs to the instability
window of the symmetry preserving solution  $E=0.5$ (red star in Fig. \ref{fig9}(a)) the system rapidly evolves to non-stationary symmetry breaking
regime. The time-Fourier power spectrum with several equidistant satellite peaks is shown in Fig \ref{fig9}(b).  It should be noticed that although there now is a pair
of stable symmetry breaking solutions at $E=0.5$ the system nevertheless does not access them but immediately transits
to the non-stationary regime. That phenomenon of satellite peak generation obviously differs from the second harmonic generation where the waves of
twice the energy (frequency) would be emitted \cite{New}.

\section{summary and discussion}

In this paper we considered the simplest nonlinear open systems whose closed analogues
allow for symmetry breaking,
namely, dimer and four-site square plaquette \cite{Eilbeck,Tsironis,Kenkre0,Kenkre,Bernstein,Tsironis1,Molina}.
The term "open" means that linear waveguides are now attached to
the nonlinear objects. The waveguides are
chosen in the form of tight-binding double chains. As shown in
Fig. \ref{fig1} this architecture preserves the mirror symmetry
with respect to the center-line of the waveguides. Then {\it if}
there are stationary solutions (\ref{stat}) the standard procedure
of matching reflected/transmitted waves (\ref{plane}) can be
applied to obtain the transmision/reflection coefficients. It is
more convenient, however, to use the Feshbach projection technique to
project the total Hilbert space onto the space of the inner states
that describe the scattering region only
\cite{Feshbach,Ingrid,Datta,SR}. The resulting equation could be
seen as a nonlinear equivalent of the Lippmann-Schwinger equation
(\ref{LS}) where $H_{eff}$ is the nonlinear non hermitian
effective Hamiltonian whose matrix elements depend in turn on the
amplitude of the injected wave. The corresponding equations are
written down for both dimer by Eq. (\ref{Datta}) and and four-site
square nonlinear plaquette Eq. (\ref{Datta4}). The effective
Hamiltonian differs from the nonlinear Hamiltonian considered in
Refs.
\cite{Eilbeck,Tsironis,Kenkre0,Kenkre,Bernstein,Tsironis1,Malomed4}
due to the presence of dissipative terms $\epsilon^2\exp(ik_p)$
where $\epsilon$ is the hopping matrix element that controls the
coupling between the closed system and the waveguides.

In case of transmission of a symmetric plane wave through the
nonlinear dimer we found two families of solutions. In the
symmetry preserving family the incident symmetric wave is
reflected and transmitted into the same symmetric channel. The second family, however, violates
the symmetry of the probing wave. It means that when a symmetric wave is injected
into the system the SB gives
rise to emission of the antisymmetric plane
waves and vice versa. Therefore the nonlinear dimer is capable
for the mode conversion, although, with maximum efficiency around 50\%. We
found that the direct solution of the time-dependent
Schr\"odinger equation with a wave front incident
to the nonlinear dimer gives the same
results for the transmission probabilities as found from the approach of the non-Hermitian Hamiltonian.
It should be pointed out that the key feature that makes possible to access the SB solutions is
the presence of domains in the parameter space where all symmetry preserving solutions are unstable.
It means that in course of time the symmetry preserving solution will eventually collapse due to the presence
of noise. The second important aspect about the open nonlinear dimer is
that symmetry is broken by both intensity and phase of the scattering function.

Similar consideration was made for the open nonlinear four-site plaquette
(see Fig. \ref{fig1} (b)). Its closed counterpart has the symmetry
group $D_4$ that provides many opportunities for the symmetry
breaking \cite{Malomed4}. However the presence of the waveguides in the
design shown in Fig. \ref{fig1} (b) substantially reduces this
symmetry to the symmetry of the open dimer. Therefore
one can expect a similar scenario for SB. However, in the case of
plaquette four nonlinear degrees of freedom participate in the transmission which
dramatically changes the dynamical picture. The standard theory of
stability \cite{Litchinitser,Cowan} based on small perturbation technique
reveals that there are domains in the parameter space where
{\it none} of the stationary solutions (neither symmetry preserving nor symmetry breaking)
are stable. It means that the scattering problem could not be reduced to stationary equations.
Direct solution of the time-dependent Schr\"odinger equation
revealed the emission of satellite waves at the energies different from the
energy of incident wave provided that this energy is chosen within the domain where the symmetry preserving
solution is unstable. The number
of satellite wave and their energies depend mostly on the
intensity of injected wave (or equivalently on the nonlinearity
constant). This effect is different from the second harmonic
generation with satellite energies not equal twice the injected wave energy.
Emergence of additional equidistant peaks in the Fourrier power spectrum of
four-cite nonlinear system was reported almost 30 years ago
in the seminal paper by Eilbeck {\it et al}
\cite{Eilbeck}. We believe that nowadays with the ongoing development of experimental
techniques, in particular in handling photonic crystal waveguides,
that phenomenon opens a new opportunity for harmonics generation. Another interesting possibility
for constructing nonlinear quantum double-chain set-ups could be Bose-Hubbard ladders in optical lattices \cite{Dhar}.

\acknowledgments{The work was supported by
Integration grant N29 from Siberian Branch of RAS and RFBR grant
N13-02-00497. The paper benefited from discussions with A.S.
Aleksandrovsky, A.R. Kolovsky, E.N. Bulgakov, K.N. Pichugin, and M.Yu. Uleysky.}

\end{document}